\documentclass[aps,pra,twocolumn,showpacs,amsmath,amssymb,preprintnumbers,superscriptaddress,10pt]{revtex4-1}

\usepackage{graphicx}
\usepackage{hyphenat}
\usepackage{multirow}
\usepackage[bookmarks=false]{hyperref}
\usepackage[capitalise]{cleveref}
\crefname{section}{Sec.}{Secs.}
\Crefname{section}{Section}{Sections}
\usepackage[dvipsnames]{xcolor}
\usepackage{makecell}

\definecolor{pink}{RGB}{255,0,255}

\begin{document}

\title{Eavesdropping and countermeasures for backflash side channel in quantum cryptography}

\author{Paulo~Vinicius~Pereira~Pinheiro}
\email{paulovpp@gmail.com}
\affiliation{Engineering Department, Para{\' i}so Faculty of Cear{\' a}, 63010-465, Juazeiro do Norte, Cear{\' a}, Brazil}
\affiliation{Department of Teleinformatic Engineering, Federal University of Cear{\' a}, C.P.6007, Campus do Pici, Fortaleza, Brazil}
\affiliation{Institute for Quantum Computing, University of Waterloo, Waterloo, ON, N2L~3G1 Canada}

\author{Poompong~Chaiwongkhot}
\email{poompong.ch@gmail.com}
\affiliation{Institute for Quantum Computing, University of Waterloo, Waterloo, ON, N2L~3G1 Canada}
\affiliation{Department of Physics and Astronomy, University of Waterloo, Waterloo, ON, N2L~3G1 Canada}

\author{Shihan~Sajeed}
\affiliation{Department of Electrical and Computer Engineering, University of Toronto, M5S~3G4, Canada}
\affiliation{Institute for Quantum Computing, University of Waterloo, Waterloo, ON, N2L~3G1 Canada}
\affiliation{\mbox{Department of Electrical and Computer Engineering, University of Waterloo, Waterloo, ON, N2L~3G1 Canada}}

\author{Rolf~T.~Horn}
\affiliation{Institute for Quantum Computing, University of Waterloo, Waterloo, ON, N2L~3G1 Canada}

\author{Jean-Philippe~Bourgoin}
\affiliation{Institute for Quantum Computing, University of Waterloo, Waterloo, ON, N2L~3G1 Canada}
\affiliation{Department of Physics and Astronomy, University of Waterloo, Waterloo, ON, N2L~3G1 Canada}

\author{Thomas~Jennewein}
\affiliation{Institute for Quantum Computing, University of Waterloo, Waterloo, ON, N2L~3G1 Canada}
\affiliation{Department of Physics and Astronomy, University of Waterloo, Waterloo, ON, N2L~3G1 Canada}
\affiliation{\mbox{Quantum Information Science Program, Canadian Institute for Advanced Research, Toronto, ON, Canada}}

\author{Norbert~L{\" u}tkenhaus}
\affiliation{Institute for Quantum Computing, University of Waterloo, Waterloo, ON, N2L~3G1 Canada}
\affiliation{Department of Physics and Astronomy, University of Waterloo, Waterloo, ON, N2L~3G1 Canada}

\author{Vadim~Makarov}
\affiliation{Russian Quantum Center and MISIS University, Moscow}
\affiliation{Department of Physics and Astronomy, University of Waterloo, Waterloo, ON, N2L~3G1 Canada}

\date{\today}

\begin{abstract}
Quantum key distribution (QKD) promises information theoretic secure key as long as the device performs as assumed in the theoretical model. One of the assumptions is an absence of information leakage about individual photon detection outcomes of the receiver unit. Here we investigate the information leakage from a QKD receiver due to photon emission caused by detection events in single-photon detectors (backflash). We test commercial silicon avalanche photodiodes and a photomultiplier tube, and find that the former emit backflashes. We study the spectral, timing and polarization characteristics of these backflash photons. We experimentally demonstrate on a free-space QKD receiver that an eavesdropper can distinguish which detector has clicked inside it, and thus acquire secret information. A set of countermeasures both in theory and on the physical devices are discussed.
\end{abstract}

\maketitle

\section{Introduction}
\label{sec:introduction}
	
Quantum key distribution (QKD) is one of the most developed branches of quantum communications. QKD offers protocol security in the sense that the QKD protocols~\cite{bennett1984,ekert1991,bennett2000} can be proven secure with a composable security definition under the model assumptions about Alice and Bob's devices, but without assumptions about the adversaries capabilities~\cite{lo1999,lutkenhaus2000,shor2000,renner2005}. Nowadays, a large number of experimental systems are available~\cite{peloso2009,peev2009,nauerth2013,idqclavis2specs,pinheiro2015,gehring2015,liao2017}. However, the issue of implementation security is still a matter of concern as the security proof of the protocols make model assumptions. By definition, there is a gap between the behaviour of actual devices and their model. This leads to possible side channels exploitable by an eavesdropper Eve~\cite{vakhitov2001,makarov2006,qi2007,lamas-linares2007,lydersen2010a,xu2010,gerhardt2011,weier2011,jouguet2013,sajeed2015,sajeed2015a,makarov2016,huang2016,sajeed2016,sajeed2017}. These side-channels may compromise the security of a QKD implementation if they are not taken into account. It is important to monitor potential side-channels and to design countermeasures to minimize their impact. With side-channels controlled in a best-practice approach, QKD will then show an implementation security that secures the generated key against future technological and algorithmic advances. The protocol security proof is an important component of that claim, even when the implementation security claim is not a mathematical proof in itself. Different aspects of QKD systems have been exploited, including but not limited to timing~\cite{zhao2008}, information leakage via Trojan-horse attack~\cite{vakhitov2001,gisin2006,jain2014,sajeed2017}, pulse-energy-monitoring system~\cite{sajeed2015,makarov2016}, device calibration~\cite{jain2011}, source flaws~\cite{xu2014}, laser seeding~\cite{sun2015}, and laser damage~\cite{makarov2016}. However, most of the reported attacks exploited detectors~\cite{makarov2006, lydersen2010a, gerhardt2011, sajeed2015a, huang2016, sajeed2016, makarov2016}, making them the most vulnerable part of the system.

Among the exploitable vulnerabilities of the detectors such as efficiency mismatch~\cite{makarov2006,zhao2008,sajeed2015a}, detector control~\cite{lydersen2010a,lydersen2010,lydersen2010b,lydersen2011,lydersen2011c,gerhardt2011,sajeed2016}, and wavelength dependency~\cite{li2011a}, one has attracted considerably less attention: the backflash emission~\cite{chynoweth1956,waldschmidt1968, childs1984, gautam1988, lacaita1993, akil1998, pacelli1998, huang2005}. It has been known for a long time~\cite{newman1955} that a reverse biased p-n junctions in a silicon avalanche photodiode in Geiger mode emits light upon the detection of a photon. Chynoweth and McKay~\cite{chynoweth1956} reported a detailed study of the phenomenon and predicted that the light emission originates due to the recombination of the energetic electrons and holes in the avalanche breakdown region. Subsequently, several other papers stated distinct possible causes for the phenomenon and quantified this emission~\cite{waldschmidt1968, childs1984, gautam1988, lacaita1993, akil1998, pacelli1998, huang2005}. In 2001, Kurtsiefer and his coworkers~\cite{kurtsiefer2001} raised the question: can this emission from the detectors employed in practical quantum communication systems affect the security? The outcome of their study suggested that the backflash photons might leak information about the detection to Eve, though the leakage of information was not quantified. Recently, a study about the backflash in InGaAs/InP avalanche photodiodes (APDs) was done~\cite{meda2016}. The latter also suggests the possibility that Eve could measure state of backflash photons and learn about detection in the receiver without causing errors in the key. 

The quantum state of the backflash photons is not expected to be correlated to that of the photon that triggered the effect. However, and unfortunately from a security point of view, the backflash photons may pass through other security critical components of Bob's receiver and carry out information about the state of those components back to the channel. For example, in polarization-based QKD with a passive basis-choice scheme, backflash photons from the horizontal (vertical) detectors will come out into the channel horizontally (vertically) polarized when they pass different arms of polarization beam-splitters (PBSes). In this case, Eve can measure the  polarization of the backflash photons and predict with high probability which detector they originated from, thus compromising the security. Another possible method of distinguishing backflash photons from different channels is monitoring the difference in time delay of backflash photons from each channel. However, for the device studied in this article, preliminary tests have shown that the difference in time delay of backflash between channels is not sufficiently distinguishable to be used to determine the source of backflash. Thus, we do not investigate the latter method here.

The Article is organized as follows. In \cref{sec:characterization_backflash}, we characterize backflash emission probability from APD and photomultiplier tube (PMT) instead of InGaAs/InP studied in Ref.~\cite{meda2016}. Furthermore, in \cref{sec:eavesdropping}, we characterize backflash photons from a free-space polarization encoding receiver, and use that information to demonstrate a practical attack on the receiver. We also quantify the information leakage to Eve in this attack scheme. In \cref{sec:countermeasure}, we introduce a countermeasure for this attack that reduces reverse transmission efficiency of the receiver from the detectors to channel to reduce information leakage. We also introduce a characterization procedure and modify the key rate equation to take into account the remaining information leakage. We conclude in \cref{sec:conclusion}.
 
\section{Characterization of backflash emission}
\label{sec:characterization_backflash}

In order to study the effect of backflash photon emission during the avalanche breakdown, a series of experiments are conducted on two different types of detectors. The first device tested is a Si-APD detector module (Excelitas SPCM-AQRH-12-FC) with a circular active area of 180~$\mathrm{\mu m}$ and peak photon detection efficiency of $0.7$ at 700~nm~\cite{excelitasAQRH}. The second device tested is a PMT (Hamamatsu H7422P-40), which has a GaAsP photocathode, with 5~mm diameter and a peak photon detection efficiency of $0.4$ at 580~nm~\cite{hamatsuh7422}. Both are thermoelectrically cooled.

\subsection{Si avalanche photodiode}
\label{subsec:apd}	

\begin{figure}[ht!]
	\centering
	\includegraphics{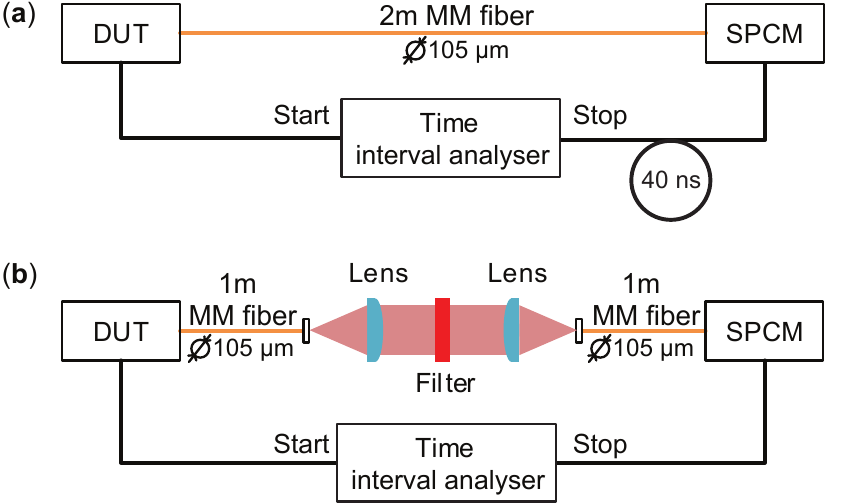}
	\caption{Setup for measuring probability of backflash emission. (a) Two identical APDs are connected with a 2~m long multimode (MM) fiber causing 10~ns optical delay between the two detectors. An electronic delay line of 40~ns is added so that the backflash photons from SPCM could also be recorded. (b) To perform spectral analysis, a free-space interference narrowpass filter is added to the setup. The filter represents one often used at the entrance of a practical QKD receiver.}
	\label{fig:setups}
\end{figure}

\begin{figure}
    \centering
	\includegraphics{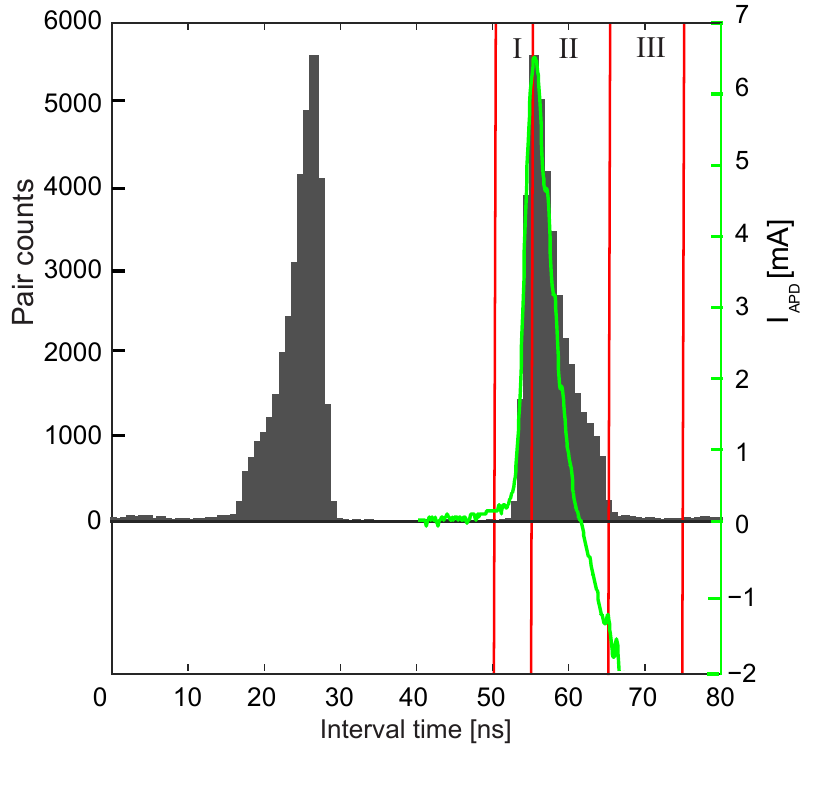}
	\caption{Histogram of time-intervals (dark grey) measured from the coincident clicks from the setup in \cref{fig:setups}. The peak on the right is backflash from DUT detected by SPCM. Regions I, II, and III of the histogram represent different stages of detector operation cycle. The shape of histogram resembles the APD current $I_\text{APD}$ (green line, measured separately). The current shape is not exact owing to a finite common-mode rejection ratio of the differential probe used to measure $I_\text{APD}$. The apparent abrupt drop of current at the border between regions II and III is common-mode interference from the quenching circuit that lowers the bias voltage and thus ends the avalanche. This coincides with a drop of photon emission almost to zero. The peak on the left is backflash from SPCM detected by DUT.}			
	\label{fig:hst1}
\end{figure}

The first step in quantifying the information leakage is to find the probability of backflash $P_b$, i.e.,\ the probability that a detection (click) leads to emission of at least one photon that leaks out of the detector. To find the value of $P_b$, we perform a measurement using the setup in \cref{fig:setups}(a). Two identical APD modules, one marked as device under test (DUT) and another marked as single-photon counting module (SPCM), are connected by a 2~m long 105~$\mathrm{\mu m}$ core diameter multimode fiber (Thorlabs M43L01). Click coincidences between them are recorded by a time interval analyzer (Stanford Research Systems SR620). In this setup, we record clicks caused by dark counts in the DUT. We record until the total clicks in DUT reach $N = 10^6$, and plot the histograms of coincidence clicks between the two detectors in \cref{fig:hst1}. The right-most peak represents the backflash photons from DUT coupled through the fiber and detected by SPCM, which occur $\approx$ 10~ns after the detections in DUT owing to the optical delay. We have added a 40~ns electrical delay so that the coincidence click appears at a delay of $\approx$ 50~ns in the plot. This also allows us to see the backflash from SPCM recorded by DUT, which is the left-most peak having a similar shape but time-inversed. The shape of the coincidence peak roughly matches that of the current flowing through the APD $I_\text{APD}$, which we have measured using a small resistor added at the APD's cathode and a wideband differential oscilloscope probe. We divide the histogram into three regions. Region I shows rapid increase in coincidence counts that resembles the exponential increase of the number of avalanche electrons flowing through the APD. Region II shows decay in the coincidence counts resembling the decrease of avalanche electrons owing to the voltage across the APD dropping as its capacitance discharges. Region III is where the voltage across the APD is further lowered below breakdown by the quenching circuit. At that time the photon emission drops to near zero. The rough match between the current shape and the photon emission suggests that the backflash photons originate from the electric current across the APD during the avalanche. 

We count coincident clicks $C$ within the right-hand peak. Here, we take into account channel transmission efficiency $T = 0.97$, and average detection efficiency of the SPCM in $500$--$900$~nm spectral band $\eta = 0.6$ \cite{excelitasAQRH}. Since the SPCM can only detect photons efficiently in this narrow spectral band, our measurement provides only a lower bound estimate of $P_b \gtrsim C/(\eta T N)$. We note that this and subsequent calculations of backflash probability are approximate in the case where $P_b \ll 1$. For this specific setup, there are 37643 coincident detections, corresponding to $P_b \gtrsim 0.065$. Furthermore, we have measured the electrical charge flowing through the APD per avalanche, by monitoring the current consumption from the high-voltage bias source. We have found that the APD under test passes on average $n_\mathrm{e^-} = 2.7 \times 10^8$ electrons through the APD per avalanche. The probability of backflash photon emission per avalanche electron $P_\mathrm{e^-} \gtrsim {P_b}/{n_\mathrm{e^-}} = 2.4 \times 10^{-10}$. We remark that a detector circuit that reduces $n_\mathrm{e^-}$ would be expected to have lower backflash.

While the wideband measurement above is imprecise, many free-space QKD setups employ a narrowband spectral filter at Bob's entrance, in order to cut background light entering Bob \cite{buttler2000,hughes2002,kurtsiefer2002,kurtsiefer2002a,bourgoin2013}. The same filter would restrict the backflash emission to the narrow band that can be measured much more precisely in our setup. We have added a free-space narrowpass filter with center wavelength of 808~nm and bandwidth of 3~nm [see \cref{fig:setups}(b)], in order to mimic spectral filter inside a practical QKD receiver~\cite{bourgoin2013}. We have repeated the counting process and found 2306 coincident detections. At this specific wavelength, the SPCM has detection efficiency of $0.62$ \cite{excelitasAQRH}. The coupling efficiency of the channel in this setup is $T = 0.83$. The probability of at least one backflash photon leaking through this filter is $P^\text{filter}_b = 4.5 \times 10^{-3}$. The spectral filter indeed reduces the emission significantly, which reduces the information leakage as we prove later in \cref{sec:countermeasure}.

We have performed another measurement to characterize the spectral distribution of the backflash photons, using a sensitive spectrum analyzer (Acton Spectrapro 2750). Unfortunately, we could not fully calibrate the spectrum analyzer for this specific setup, and the result is only qualitative. The measurement indicates that the backflash emission is broadband, spanning continuously from 550~nm to $>$1000~nm with a gentle peak around 900~nm (see \cref{sec:spectrum}). This broadband characteristic leads to the possibility of including a narrow bandpass spectral filter in the system. The filter limits the wavelength range in which the backflash probability needs to be characterised, reduces the backflash emission from Bob, and thus reduces the information leakage. 

\subsection{Photomultiplier tube}
\label{subsec:pmt}
		
Photomultiplier tube (PMT) is another type of detector widely used for its larger sensitive area and moderate dark count rate~\cite{hadfield2009}. We have replaced DUT in \cref{fig:setups}~(a) with a PMT unit. Since the dark count rate of the PMT is low, additional weak laser pulses have been coupled to the active area of PMT to induce clicks. After recording $10^6$ counts in the PMT, we have found fewer than 100 coincidences for both the fiber and free-space setups. This coincidence level is close to the dark count level of the SPCM, implying that the probability of backflash in PMT is negligible within the spectral range of our measurement.

\section{Eavesdropping experiment}
\label{sec:eavesdropping}

\begin{figure}
    \centering
	\includegraphics{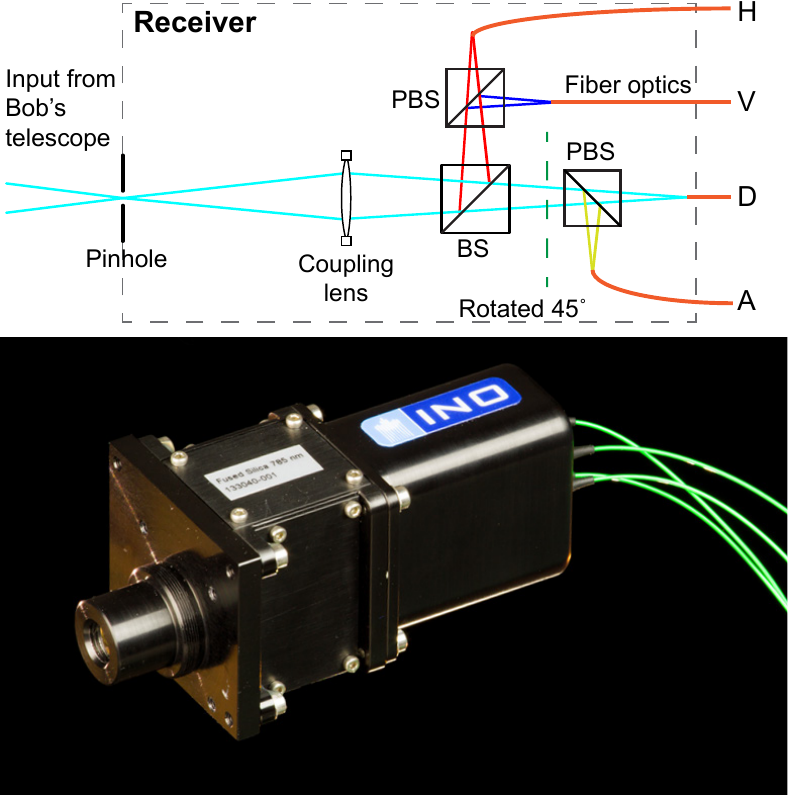}
	\caption{Receiver designed by INO working as a passive basis choice polarization analyzer at 785~nm. Top: the important optical components consists of a pinhole, coupling lens, beamsplitter (BS), and polarizing beamsplitters (PBSes). Bottom: photo of the receiver. Four multimode fibers lead to the four detectors (not shown).}
	\label{fig:receiver}
\end{figure}

In this section, we experimentally quantify Eve's ability to identify which detector the backflash photons originated from, by measuring the backflash photon's polarization state. Bob's receiver used in this test is an integrated receiver built by INO (National Optics Institute of Canada) designed for a free-space passive polarization encoding QKD system running at 785~nm. \cref{fig:receiver} shows its optical scheme. The receiver consists of a pinhole to prevent spatial mode attack~\cite{sajeed2015a}, coupling lens to focus incoming beam into optical fibers, and an integrated optics module. The latter consists of a beamsplitter (BS) to passively select the basis of measurement and PBSes in each basis to discriminate the four polarizations of the incoming photons: horizontal (H), vertical (V), diagonal (D), and antidiagonal (A). Next, we characterize the backflash emission as a possible side channel.

\subsection{Reverse loss and extinction ratio}
\label{subsec:loss}

\begin{figure}
    \centering
	\includegraphics{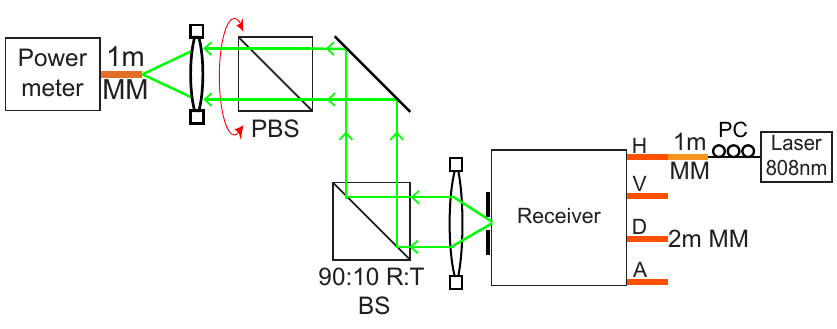}
	\caption{Setup for measurement of the reverse propagation loss and polarization extinction ratio. An 808~nm laser is connected to each of the output channels of the receiver, one at a time. A 90:10 reflection:transmission (R:T) ratio beamsplitter diverts the reverse propagating beam to the measurement unit. The latter consists of a fiber-coupled optical power meter, and a rotating PBS to measure power and polarization extinction ratio of the reverse propagation beam. A polarization controller PC is used to maximize throughput power from each receiver channel.}
	\label{fig:setup_loss}
\end{figure}

As the photons back-propagate through the setup, they experience the reverse loss of the receiver, i.e.,\ the loss from originating detector to the channel input. This could reduce probability that backflash photon leaks into the channel. The setup shown in \cref{fig:setup_loss} is used to estimate this loss. An 808~nm laser (wavelength close to the operating wavelength of the receiver) is connected to the receiver's output multimode fiber, one channel at a time. The laser power at the end of receiver's fiber is $P_1 = 40$~$\mathrm{\mu W}$. We adjust the polarization controller PC to maximize throughput power, providing an upper bound of the reverse transmission. We then measure laser power $P_2$ emitted at the front of the receiver module, between the focusing lens and receiver's pinhole in \ref{fig:setup_loss}. The reverse transmission efficiency of the receiver for the optimum polarization is then $T_b = P_2/P_1$. We have measured the average reverse transmission efficiency over all four channels of this receiver $T_b\approx 0.091$ (the individual values lie in the range $0.088$ to $0.094$). Assuming backflash photons are randomly polarized, their transmission should be approximately half of this upper bound.

\begin{table}
	\centering
	\caption{\textbf{Reverse propagating extinction ratio measurement of Bob's setup. The photons from H and V channel could be distinguished with high probability. The measured extinction ratios of A and D channels are low, presumably owing to polarization becoming elliptical at reflections in the measurement unit.}}
	\label{tab:system_loss}
	\renewcommand{\arraystretch}{1.2}
	\begin{tabular}{c c c c c c}
	\hline
	\multirow{2}{*}{\makecell{Output\\ channel}} & \multicolumn{2}{c}{max} & \multicolumn{2}{c}{min} & \multirow{2}{*}{\makecell{Extinction\\ ratio}} \\ \cline{2-5} & {\makecell{Angle\\ (deg)}} & {\makecell{Power\\ ($\mathrm{\mu W}$)}} & {\makecell{Angle\\ (deg)}} & {\makecell{Power\\ ($\mathrm{\mu W}$)}} & \\
	\hline
	H      & 3      & 25.0    & 91    & 0.15    & 167    \\ 
	V      & 94     & 19.8    & 1     & 0.03    & 660    \\ 
	D      & 315    & 20.7    & 223   & 1.94    & 10.7   \\ 
	A      & 49     & 23.5    & 141   & 3.69    & 6.4    \\
	\hline
	\end{tabular}
\end{table}

Next, we demonstrate Eve's ability to distinguish the originating channel of backflash photon. For that, we measure polarization extinction ratio of the reverse emitted beam from the receiver. In \cref{fig:setup_loss}, a 90:10 reflection:transmission (R:T) ratio beamsplitter is added to divert the outgoing beam from the receiver to a measurement unit consisting of a PBS and a fiber-coupled optical power meter. This additional setup has throughput efficiency $T_e = 0.60$. For each receiver channel input, we rotate the PBS to find a pair of angles that results in maximum and minimum power at the power meter. The optimal angles for each channel and respective extinction ratios are shown in \cref{tab:system_loss}. The drastically lower extinction ratio in D and A polarization is likely a result of polarization distortion caused by Fresnel effect on the dielectric mirror and the 90:10 BS used by Eve. These reflective surfaces were aligned at a certain angle along the axis corresponding to V polarization. This alignment distorted the diagonal polarization of the reflected beam, by inducing a phase difference between its H and V polarization components. In real eavesdropping, Eve can correct this polarization distortion using a phase compensator or waveplate. She can also split the incoming backflash photons into two PBSes oriented at the angles that yield the highest extinction ratios in both bases. This should allow her to distinguish the photons from all four channels with high probability.

\subsection{Timing of backflash photons through the receiver}
\label{subsec:time}
		
The previous experiment suggests that by measuring the polarization of the backflash photons, Eve could estimate which detector they originated from. However, in real life scenario, Eve's detection might not solely be from the backflash photons; it can be a result of stray light in the channel, reflection of Alice's signal from Bob's optical components or dark counts in Eve's detector -- all unwanted noise. To avoid those unwanted signals, Eve needs to synchronize her measurement apparatus with Alice and Bob's signal pulses, and activate her detector at a specific time when the backflash photons are expected to arrive. The synchronization can be done by monitoring Alice's and Bob's signals prior to the eavesdropping. This section demonstrates a practical setup to measure timing characteristics of the backflash photons. 

\begin{figure}
	\centering
	\includegraphics{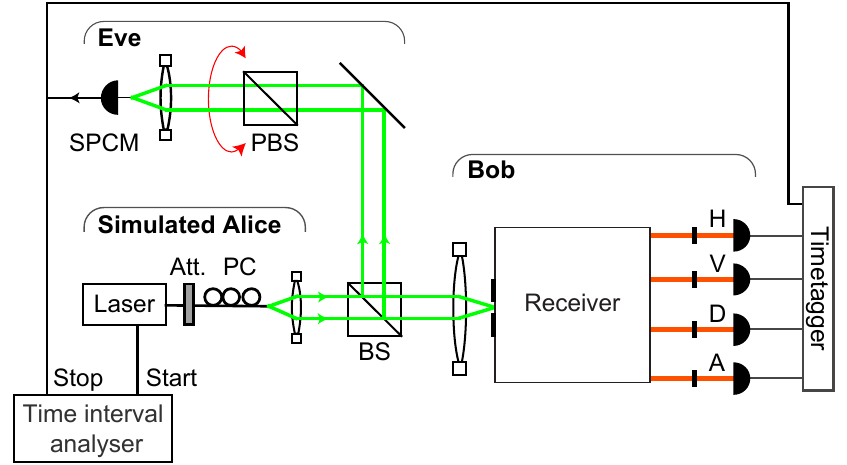}
	\caption{Eavesdropping setup for timing characterization and proof-of-principle attack. The 90:10 R:T BS diverts photons from Bob to Eve's detector. Eve's setup consists of a PBS that can be rotated to find the optimal angle for Eve to distinguish the source of backflash photon. The time interval analyser (TIA) is used to find the time delay of the backflash photon in the channel. The timetagging unit records coincidence time between Bob's and Eve's detections in the proof-of-principle attack.}
	\label{fig:setup_timetag}
\end{figure}

\begin{figure}
    \centering
	\includegraphics{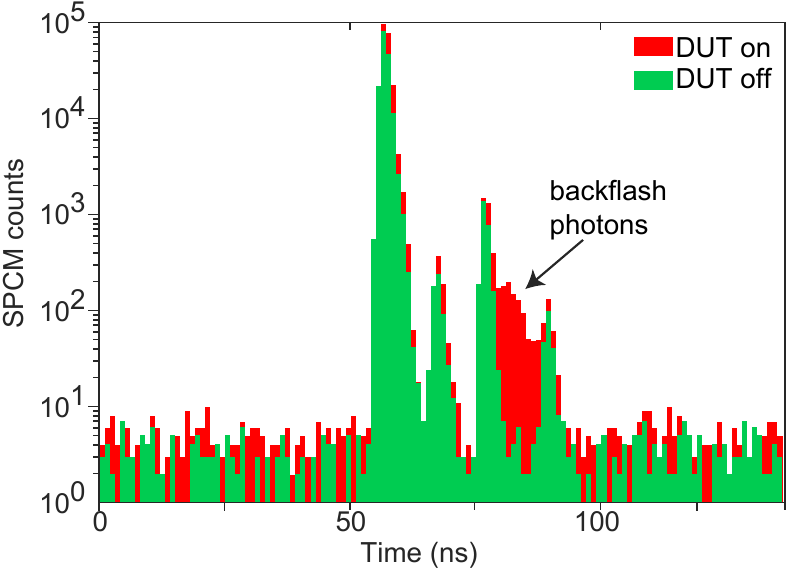}
	\caption{Histogram of time intervals between emitting Alice's laser pulse and detection in Eve's SPCM. The histogram with DUT powered on (red) has an area of coincidence peak well above the level when DUT is powered off (green). The timing of this area matches the optical time delay between Eve's receiver and DUT, indicating backflash emission. The other peaks are optical reflections in the setup (see text for details).} 
	\label{fig:pulse}
\end{figure}

The experimental setup is shown in \cref{fig:setup_timetag}. A train of $3~ns$ wide laser pulses with $200~ns$ period is sent to Bob's receiver to simulate signals from Alice. The detector used as DUT in \cref{subsec:apd} is connected to one channel of the receiver at a time. A time interval analyzer (TIA) is used to record the coincidence time between the signal sent by Alice and Eve's SPCM clicks. In \cref{fig:pulse}, we plot two histograms of the coincidence time from the APD in H channel. The green histogram is the coincidence time when DUT is powered off. Thus the detections in Eve resulted from reflections from the receiver's optical components. The positions of the peaks correspond to optical delay between reflective components in the setup and Eve's SPCM. The leftmost peak is a result of backreflection off the free-space optics at the front of Bob, such as his lenses and BS. The next peak matches the time delay from fiber splices in the receiver's fiber, indicated by short bars in \cref{fig:setup_timetag}. The third peak is the backreflection from the APD (in H channel only, as the fiber in the other channels has been terminated with matching gel that eliminates backreflections). The time delay of the right-most peak matches the round-trip of triple reflection between the APD and fiber splice. The red histogram is the coincidence time when DUT is powered on. Extra counts due to backflash photons can clearly be seen at $80$--$87~ns$. The time delay matches optical delay between DUT and Eve's SPCM. Since the coincident counts of backflash events are $\approx 1.5$ orders of magnitude higher than the back-reflection and noise level, the probability of Eve registering back-reflected pulses within this time window is small. Similar result could be seen when connecting the DUT to V, D, and A channels.

\subsection{Proof-of-principle eavesdropping demonstration}
\label{subsec:attack}

We next emphasize the threat of this attack by demonstrating Eve's performance using a practical setup, shown in \cref{fig:setup_timetag}. In this experiment, we demonstrate Eve's ability to distinguish backflash emissions in one basis, between H and V channels. We only consider those photons that are coupled back to the optical channel and thus could carry information to Eve. We first repeat the alignment procedure as described in \cref{subsec:loss} by sending laser beam through the receiver's fibers, and rotating the PBS in Eve to find two optimal angles where the detection rate from the laser sent through Bob's H channel is maximum but V channel is minimum, and vise versa. Bob is then equipped with four powered-on APDs, one at each channel of the receiver, as in a real QKD setup. As seen in \cref{subsec:time}, Eve needs to register the coincidence counts within a specific time window to filter out back-reflection events. For that, we replace TIA with a timetagger (Dotfast Consulting 78-ps resolution 8-channel module) set to register the events where Eve's detector clicks within 25--30$~ns$ after Bob's detection, which matches the time delay between Bob's and Eve's detectors. A train of $3~ns$ wide laser pulses with $200~ns$ period are sent to Bob to simulate QKD signal pulses from Alice. For each orientation of Eve's PBS, we count the number of detections in Bob and coincidence count in Eve over $10~s$. We record the ratio of coincidence events $R_{ij} = E_{ij}/B_i$, where $B_i$ is the number of clicks in Bob's $i$th detector, and $E_{ij}$ is the number of Eve's coincident clicks with Bob's $i$th detector when she sets her PBS angle to maximise clicks from Bob's channel $j$. For example, $R_{HH}$ represents probability of a click in Bob's H channel causing a coincident click while Eve aligns her PBS to measure signal from H channel, i.e., the probability that Eve gets a correct detection. 

The probability of Eve gaining information (about H channel detection) is the chance of getting correct detection ($R_{HH}$) less the chance that she gets a wrong detection ($R_{HV}$). Note that the backflash probability $P_b$ and reverse transmission efficiency $T_b$ are already accounted in these coincidence ratios. Our measurements show that, for Bob's H detection, $R_{HH} = 5.00 \times 10^{-3}$ and $R_{HV} = 1.45 \times 10^{-3}$, causing information leakage of $3.5 \times 10^{-3}$. For Bob's V detection, $R_{VV} = 5.69 \times 10^{-3}$ and $R_{VH} = 3.66 \times 10^{-3}$, causing information leakage of $2.0 \times 10^{-3}$. From the calibration measurements we have expected the information leakage to be less than $\eta T_e T_b P_b/2 = 1.1 \times 10^{-3}$, which poorly matches the leakage observed in the eavesdropping experiment. We could not explain this discrepancy.

This result shows that Eve could learn a fraction of Bob's detections by monitoring the backflash photons. On the one hand, the information leakage is small, and we don't have the spectral filter in Bob in this experiment. On the other hand, our Eve's setup is not an optimal one for the attack. Proper countermeasures both in physical implementation and in post-processing step need to be considered.

\section{Countermeasure}
\label{sec:countermeasure}

In this section, we discuss about possible countermeasures for attacks exploiting backflash photons. For physical implementation, using PMT can eliminate the possibility of generating backflash photons (although this conclusion is subject to the limitations of our measurement in \cref{subsec:pmt}). 
Another possible countermeasure is using measurement-device-independent
QKD (MDI-QKD) \cite{lo2012, liu2013}, in which the detection outcomes are public, thus
Eve gains no new information from the backflash. However implementation of
MDI-QKD in free-space is challenging \cite{qi2015, wang2015, diamanti2016}.
If a non-MDI-QKD system uses APDs,
the information leakage could be limited by decreasing reverse transmission efficiency $T_b$ either by adding narrow-band spectral filter as shown in \cref{sec:characterization_backflash}, or an optical isolator. These measures could reduce but not eliminate the leakage of information. The remaining leakage needs to be taken into account when calculating the required shortening of the key during privacy amplification.

The following procedure could be employed. Bob follows the procedure in \cref{subsec:apd} to find the APD's probability of backflash $P_b$ and receiver's reverse transmission efficiency $T_b$. This $T_b$ includes all optical isolators and filters added to the receiver to limit the information leakage. If Bob does not include a narrow-pass filter, these parameters need to be characterized in a very wide spectral range, because typical free-space optics and air are transparent in a wide spectral band. This wide spectral characterization will be challenging. However, if a band-pass filter is used, it is sufficient to characterize the parameters over its spectral pass-band. From the result in \cref{subsec:loss}, it is reasonable to assume that in the worst case, with ideal equipment, Eve could distinguish the origin of backflash photons with certainty. The information leakage to Eve is then $P_E = P_b T_b$. In other words, a fraction $P_E$ of Bob's detections is tagged by Eve without disturbing the quantum state or inducing error. Then the privacy amplification for QKD with tagged signal~\cite{gottesman2004,lutkenhaus2000} can be used to take care of the information leakage.

As an example, let us consider the key rate equation for the Bennett-Brassard 1984 (BB84) protocol in QKD system with single-photon signals. Under the backflash attack, the secret key rate per signal sent by Alice becomes
\begin{equation}
l \geq AP_{det}(1-h(\frac{e}{A}))-leak_{EC},
\end{equation}
where $P_{det}$ is the probability of detection per signal, $e$ is the error rate, $h(x) = -x\log{x}-(1-x)\log(1-x)$ is the binary Shannon entropy, and $leak_{EC}$ is the portion of key disclosed during error correction. The correction term $A = (P_{det}-P_E)/P_{det}$, where $P_E$ is the information leakage calculated in the characterization step above. 

The theoretical analysis in this paper considers only the worst-case scenario where Eve has the ability to collect and distinguish all backflash photons and map them to the raw key in Alice and Bob. This analysis also provides only the lower bound on the secret key rate, which could be improved by more careful analysis.

\section{Conclusion}
\label{sec:conclusion}

We have quantified the backflash emission of photons from APD-based single-photon detectors, and verified that these photons can be used by an eavesdropper to learn about the key in QKD systems. We have found that, for a system without spectral filter, at least $0.065$ of the clicks in actively-quenched Si detector module result in backflash. This probability is reduced by a factor of 14 when a narrowband spectral filter is added, suggesting the latter is an efficient countermeasure. For PMT the backflash emission is negligible within the sensitivity of our measurement. Our experiment with a real polarization-encoding QKD receiver shows that Eve can distinguish polarization of backflash photons with near certainty. The proof-of-principle attack shows that Eve could learn $2.0 \times 10^{-3}$ fraction of raw key using our today's imperfect setup. The information leakage may be higher for an ideal Eve. To close this loophole, we discuss a procedure to characterize the system and quantify Eve's information, then modify the key rate equation to take care of the information leakage due to backflash emission. We hope that our study will contribute to the development of certification and standardization of practical QKD against side-channels.

\appendix
\section{Spectral distribution measurement}
\label{sec:spectrum}

\begin{figure}[h]
\centering
	\includegraphics{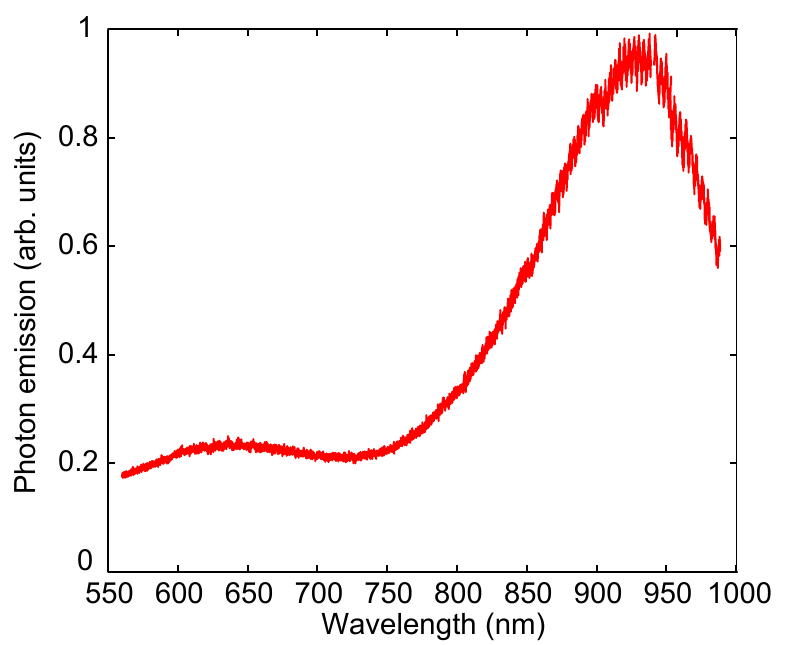}
	\caption{Spectral distribution of backflash.}			
	\label{fig:spectrum}
\end{figure}

\Cref{fig:spectrum} shows the spectral distribution of backflash emission measured with a sensitive spectrum analyzer (Acton Spectrapro 2750). Due to difficulties we have encountered in spectrometer calibration, this measurement has a large margin of error comparing with the narrow-band filter measurement at a specific wavelength. Thus, we omit this result from the main Article. Even so, this measurement shows some important characteristics of backflash emission. The backflash emission is broadband, spanning continuously across our range of measurement from $550$ to $1000$~nm with a gentle peak around 900~nm. This suggest the possibility of having backflash emission beyond our range of measurement. This emphasizes the necessity of adding the narrow-band filter to ease the characterization process and limit the information leakage.

\section*{Acknowledgments}
This work was supported by the US Office of Naval Research, Industry Canada, CFI, Ontario MRI, NSERC (programs Discovery and CryptoWorks21), and the Canadian Space Agency. P.V.P.P. was also supported by a Brazilian CAPES scholarship (project no.\ 014633/2013-02). P.C.\ was also supported by a Thai DPST scholarship. J.-P.B.\ was also supported by FED~DEV.

Author contributions: P.V.P.P.\ and P.C.\ contributed equally to this study. P.V.P.P.\ conducted measurements on the individual detectors and integrated receiver. P.C.\ conducted the eavesdropping demonstration experiment and theoretical analysis. S.S.,\ R.T.H.,\ and J.-P.B.\ assisted in setting up the experiments. N.L.\ supervised the theoretical analysis. V.M.,\ T.J.,\ and N.L.\ supervised the study. P.C.\ and P.V.P.P.\ wrote the article with input from all authors.

\def\bibsection{\medskip\begin{center}\rule{0.5\columnwidth}{.8pt}\end{center}\medskip} 
\bibliography{bibtex_library}
\bibliographystyle{plain}
\end{document}